# Ultrafast Transverse Modulation of Free Electrons by Interaction with Shaped Optical Fields

I. Madan[1], V. Leccese[1], A. Mazur[2], F. Barantani[1,3], T. La Grange[1],
A. Sapozhnik[1], S. Gargiulo[1], E. Rotunno[4], J.-C. Olaya[2], I. Kaminer[5],
V. Grillo[4], F. J. García de Abajo[6,7], F. Carbone[1,*], and G. M. Vanacore[8,*]

1. Institute of Physics, École Polytechnique Fédérale de Lausanne, Lausanne, 1015, Switzerland
2. HOLOEYE Photonics AG, Volmerstrasse 1, 12489 Berlin, Germany
3. Department of Quantum Matter Physics, University of Geneva, 1211 Geneva, Switzerland
4. Centro S3, Istituto di Nanoscienze-CNR, 41125 Modena, Italy
5. Department of Electrical and Computer Engineering, Technion, Haifa 32000, Israel
6. ICFO-Institut de Ciencies Fotoniques, The Barcelona Institute of Science and Technology, 08860 Castelldefels (Barcelona), Spain
7. ICREA-Institució Catalana de Recerca i Estudis Avançats, Passeig Lluís Companys 23, 08010 Barcelona, Spain
8. Department of Materials Science, University of Milano-Bicocca, Via Cozzi 55, 20126 Milano, Italy

*To whom correspondence should be addressed: giovanni.vanacore@unimib.it, fabrizio.carbone@epfl.ch

**Spatio-temporal shaping of electron beams is a bold frontier in electron microscopy, enabling new routes toward spatial-resolution enhancement, selective probing, low-dose imaging and faster data acquisition. Over the last decade, shaping methods evolved from passive phase plates to low-speed electrostatic and magnetostatic displays. Recently, higher shaping speed and flexibility have become feasible by the advent of ultrafast electron microscopy, embodying a swift change of paradigm that relies on using light to control free electrons. Here, we experimentally demonstrate that arbitrary transverse modulation of electron beams is possible without the need for designing and fabricating complicated electron-optics elements or material nanostructures, but rather resorting to shaping light beams reflected from a planar thin film. We demonstrate arbitrary transverse modulation of electron wavepackets via inelastic interaction with a shaped ultrafast light field controlled by an external spatial light modulator (SLM). We illustrate this method by generating Hermite-Gaussian (HG) electron beams with $HG_{10}$ and $HG_{01}$ symmetry and discuss their possible use in enhancing the imaging contrast of microscopic features. Relative to current schemes, our approach adopting an external SLM dramatically widens the range of patterns that can be imprinted on the electron wave function and makes electron shaping a much easier task to perform.**

For a long time, electron microscopy has epitomized the art of acquiring images at increasingly higher spatial resolution [1]. Instrumentation research was mainly aimed at obtaining atomic-size probes and aberration-free images. About ten years ago a different trend was started when the first ideas of electron beam manipulation were introduced [2,3,4,5]. These methods mainly relied on the use of phase and amplitude holograms [6,7,8,9,10,11,12], as well as electrostatic and magnetostatic phase elements [13,14,15,16,17], to coherently modulate the amplitude and phase of the transmitted free-electron wave function [18,19,20,21]. Thanks to such advances, beam shaping can now provide new routes toward image-resolution enhancement, selective probing, low-dose imaging, depth information, as well as faster data acquisition [8,22,23]. We anticipate that such advancements will be able to open new frontiers not only in microscopy but also in optoelectronics, quantum information and biosensing.

To reach the required high speed, flexibility and precise phase control needed to transform this strong potential into a reality, a radical departure from current passive or slowly-varying schemes becomes necessary. A swift change of paradigm has recently occurred with the exploration of new approaches based on light-mediated coherent modulation of the longitudinal [24,25,26,27,28,29,30] and transverse [24,31,32,33,34,35] amplitude and phase of an electron wave function [36,37,38,39,40]. Such schemes exploit the effect of strong interaction between free electrons and electromagnetic fields taking place either in free-space through elastic coupling mediated by the ponderomotive force of a standing wave of light [41,42,43] (only phase modulation), or in presence of fabricated nanostructures via inelastic exchange of photon quanta [44,45,46,47,48,49,50] (both phase and amplitude modulations when using energy filtering). The use of properly synthesized ultrafast light fields can provide coherent shaping of the electron wave with temporal modulation speed down to the femtosecond range and below, many orders of magnitude faster than in conventional electrically-controlled schemes.

Here, we present the first step towards the development of such a Photonic free-ELectron Modulator (PELM) and experimentally demonstrate the transverse modulation of the electron wave function with a computer-controlled arbitrarily-shaped transverse profile of the ultrafast light field. In our approach, which is schematically shown in Fig. 1a, instead of using fabricated nanostructures to create a specific light field configuration, we adopt an external spatial light modulator (SLM) to imprint the desired amplitude and phase pattern on the optical field, which is in turn projected on a flat electron-transparent plate. Such a pattern is then imprinted on the electron wavepacket when it crosses the light field via inelastic electron-light scattering. The resulting electron pattern is then observed by energy-filtered ultrafast electron microscopy [51,52]. Using an SLM thus enormously widens the range of light patterns that can be adopted for electron beam shaping with respect to the limited set of configurations defined by specific fabricated nanostructures. By precisely tuning the phase and intensity of the light field via the SLM, such a method allows us to externally and arbitrarily manipulate the transverse and longitudinal electron distribution with an unprecedentedly-high control of electron beams with programmable time/energy and space/momentum distributions.

In essence, we overcome the problem of designing and fabricating complicated electron-optics elements by resorting to shaping light beams, which has been proven a much easier task to perform. At the same time, the method allows an unprecedented access to fast temporal modulation. Indeed, a critical advantage of our approach with respect to existing technologies lies in the capability to achieve ultrafast switching of the electron wave profile and an extremely flexible electron manipulation. The fast, tailored, and versatile control achievable with such ultrafast light fields would allow us to simultaneously engineer the spatial, temporal, spectral, and momentum distributions of an electron in a coherent manner, providing new approaches for the investigation of ultrafast excitations in materials [53,54,55,56,57,58,59,60,61].

As shown in Fig. 1a-b, a suitable platform for generating the required light field configuration is provided by a tilted light-opaque, electron-transparent thin film in which an externally controlled optical pattern is projected from a SLM placed in the conjugate plane with respect to the film surface. In our design, we adopt a 30-nm thick Ag layer deposited on a 20-nm thick $Si_3N_4$ membrane. Femtosecond electron pulses impinge on the $Ag/Si_3N_4$ plate and interact with the semi-infinite light field created at the Ag surface *via* stimulated inverse transition radiation [26]. The inelastically-scattered electrons are then imaged in space and time by electron energy-loss spectroscopy (EELS) (see the Methods section for further experimental details).

In Fig. 1c, we show EELS spectra recorded as a function of the delay time between electron and light pulses. Following the interaction, the zero-loss peak (ZLP) at an electron energy $E_0$ = 200 keV is redistributed among sidebands at multiples of the incident photon energy $\pm \ell \hbar \omega$, corresponding to energy losses and gains by the electrons of $\ell$ photon quanta (in our case $\hbar\omega = 1.57$ eV). The electron-light interaction is captured by a single complex coupling coefficient [26,62]

$$\beta_{PELM}(x,y) = (e/\hbar\omega) \int dz'\, \mathcal{E}_z^{SLM}(x,y,z')e^{-i\omega z'/v}. \tag{1}$$

where $v$ is the electron speed and $e$ is the elementary charge. The parameter $\beta$ depends on the component of the light electric field $\mathcal{E}_z^{SLM}$ along the electron propagation direction and on its distribution over the transverse (in-plane) coordinates $(x,y)$. We remark that the transverse dependence is key for our photonic electron modulator.

Following the interaction with the optical field, the initial electron wave function $\psi_0$ gains inelastic components labeled by the net number of photon exchanges $\ell$ according to

$$\psi_{PELM,\ell}(x,y) = \psi_0 J_\ell(2|\beta_{PELM}|)e^{i\ell \arg\{-\beta_{PELM}\}}, \tag{2}$$

where $J_\ell$ is the $\ell^{th}$-order Bessel function of the first kind.

In our approach, we use the SLM to imprint on the optical field a transverse amplitude and phase pattern, which is then directly transferred to $\beta(x,y)$ and, thus, to the amplitude and phase of the inelastic electron wave components $\psi_\ell(x,y)$. This is demonstrated here for the case of Hermite-Gaussian beams.

In Fig. 2a-c, we show the phase patterns implemented on the SLM. When using a homogeneous phase distribution, the SLM behaves as a regular homogeneous mirror, and the gaussian light pulses are thus focused unperturbed on the $Ag/Si_3N_4$ thin film (see Fig. 2d). Instead, the introduction of a well-defined π phase shift in half of the SLM, separated by a boundary along either the horizontal direction (Fig. 2b) or the vertical direction (Fig. 2c), induces the formation of Hermite-Gaussian (HG) beams in the conjugate plane with $HG_{10}$ (Fig. 2e) or $HG_{01}$ (Fig. 2f) symmetry, respectively.

In Fig. 2g-j, we show the inelastically-scattered electron spatial maps – see Methods for details – measured under optical illumination with Gaussian and Hermite-Gaussian beams obtained by SLM modulation of the ultrafast light pulses. The images are taken when the electron and the light wavepackets are in temporal coincidence. Here, we immediately notice that the characteristic Hermite-Gaussian distributions are directly imprinted on the transverse profile of the electron pulse, which changes from a $HG_{10}$ (Fig. 2h) to a $HG_{01}$ (Fig. 2j) symmetry according to the in-plane pattern on the optical field.

As described in Ref. 24, the coupling coefficient $\beta$ for the adopted experimental geometry of a semi-infinite light field reflected from a 30-nm thick Ag layer (treated as a perfect-conductor mirror) is given by

$$\beta^{\dagger}_{PELM}(x,y) \approx \frac{ie}{\hbar\omega}\left[\frac{\mathcal{E}_z^{inc}(x,y)}{\omega/v - k_z^{inc}} + \frac{\mathcal{E}_z^{ref}(x,y)}{\omega/v - k_z^{ref}}\right], \tag{3}$$

where $\mathcal{E}_z^{inc/ref}$ and $k_z^{inc/ref}$ are the projections of the incident/reflected SLM light electric field and wave vector, respectively, on the *z* direction defined by the electron wave vector, which coincide with the plate surface normal in the present instance. These quantities depend on the tilting geometry of the layer with respect to the electron and light directions. Their detailed expressions for the geometrical parameters used in the experiments are offered in Supplementary Section 1. In Fig. 3d-f, we plot the calculated incident transverse profiles of the light pulse used for the experiments performed in this work, where either a Gaussian or a Hermite-Gaussian beam is made to interact with the electron pulse. Upon propagation of the electron wavepacket through such field configuration, the inelastically-scattered electron spatial maps can be obtained as

$$I_{PELM}(x,y) = \sum_{\ell=-1}^{-\infty} |\psi_{PELM,\ell}(x,y)|^2 \approx |\psi_0|^2 \sum_{\ell=-1}^{-\infty} \left|J_\ell\big(2|\beta^{\dagger}_{PELM}(x,y)|\big)\right|^2, \tag{4}$$

where the summation runs over the energy gain components of the electron wave (i.e., $\ell < 0$). The calculated maps are shown in Fig. 3g–j and correctly reproduce the two-lobed probability distributions observed in experiment, as induced by the SLM phase modulation of the optical field. It is worth mentioning that the tilt of the electron lobes is a result of the off-normal light incidence on the Ag mirror.

Having demonstrated the ability to generate ultrashort Hermite-Gaussian femtosecond electron beams, we discuss their possible application in the dynamical investigation of materials. Because HG beams are the electron equivalent of linearly polarized light, they are particularly sensitive to the mode symmetry of localized electromagnetic fields [63,64]. For instance, ultrafast HG beams with a well-defined transverse phase coherence and azimuthal order can act as selective probes of the dynamical behavior of specific plasmonic resonances characterized by the corresponding charge multipolarity. These types of probes hold an enormous potential for the investigation of the spatio-temporal field evolution in photonic cavities and metamaterials, where multiple degenerate modes of different symmetry are usually excited at unison.

Additionally, shaped electron beams can also be used to enhance the contrast in TEM images from weak scatterers [65,66,67,68,69]. This can lead either to a reduction of the total electron dose needed to form an image – thus reducing the radiation damage – or to an improvement (with the same electron dose) of the signal-to-noise-ratio in such image. The latter aspect would be beneficial for the investigation of the real-time evolution of beam-sensitive samples in their natural environment.

As an example of application of Hermite-Gaussian modulation in TEM imaging, we demonstrate the possibility of using HG beams to enhance the topographic contrast of an ordered array of Skyrmions in the limit of the weak-phase approximation. In terms of electron shaping schemes, the Hermite-Gaussian modulation of the electron wavepacket induced by the PELM is equivalent to the application of a Hilbert-phase plate (HPP) with the corresponding symmetry. The HPP, which imposes a π phase shift in half of the interaction region, is a well-known and well-characterized type of phase plate that is used in imaging of weak-phase objects [65,66,67,68]. For illustration, we consider a hexagonal array of Skyrmions (diameter of ~ 50 nm) separated by a distance of 100 nm and perform simulations using the STEM-CELL software [70]. TEM imaging of Skyrmions is generally performed in Lorentz microscopy in Fresnel mode with a defocus of hundreds of microns or more [71,72]. In Figs. 4a and 4b, we present the simulated TEM images of the Skyrmions lattice obtained at the focus condition (zero defocus) with Hermite-Gaussian electron beams by the application of two orthogonal HPPs having the corresponding symmetries, respectively. In Fig. 4c, we show the quadratic average of the two HG images. The latter provides a highly-resolved image of the Skyrmions lattice with a strong increase of the local contrast with respect to conventional LTEM images, where lateral resolution is limited by the large defocus. Such a method, which takes advantage of the unique possibility of PELM to rapidly alternate between the two symmetries of the HG beam, could be used as starting point for quantitatively reconstructing the phase of the transmitted electron-beam wave function following the interaction with the sample.

The approach demonstrated here is, of course, significantly more general than just the generation of Hermite-Gaussian electron beams shown above. Our results represent the first step toward the realization of an all-optical rapidly programmable phase mask for electrons. The inelastic interaction between an electron pulse and an arbitrarily-shaped ultrafast light field controlled by an external spatial light modulator allows us to achieve ultimate amplitude and phase control of the electron wavepacket in space and time. Besides the fundamental aspects, this will have a strong technological impact in enabling new imaging methods in electron microscopy with enhanced performances – especially in terms of sensitivity and resolution, which are hardly attainable based on conventional static or slowly-varying schemes. We thus anticipate our approach to be a step forward in the ability to radically change how matter is investigated in electron microscopy.

As an appealing possibility, optically-induced transverse amplitude modulation of electron wavepackets would be ideal for the implementation of an electron single-pixel imaging (ESPI) scheme in space and time [61], where we would use laterally-structured electron pulses to illuminate the object of interest while synchronously measuring the total intensity of the inelastically-scattered electrons. In a TEM, spatial and temporal SPI has never been implemented so far, mainly due to the lack of fast and versatile electron modulators capable of generating the required rapidly changing electron patterns. Therefore, our results are extremely promising in terms of the experimental realization of Electron-SPI, which will enable microscopic investigations with lower noise, faster response time, and lower radiation dose with respect to conventional imaging approaches.

An additional intriguing opportunity opened by the PELM design is the generation of high-purity ultrashort vortex electron pulses with a designated azimuthal order *via* interaction with vortex light pulses created by the SLM pattern. Such configuration will allow us to implement time-resolved electron chiral dichroism, enabling ultrafast chiral sensitivity at the ultimate temporal and spatial resolution. In comparison with conventional methods, the key aspect of our approach would be the possibility to time-lock the SLM with the laser and the electron detector, providing fast vorticity switching of the electron pulse and homodyne extraction of the dichroic signal. This approach will thus provide a tool able to access the out-of-equilibrium behavior of chiral excitations in quantum materials – one of the latest challenges in condensed-matter physics – such as chiral phonons in 2D van der Waals layers with broken space-inversion symmetry [73] or chiral plasmons in Dirac systems [74].

These new imaging methods enabled by the arbitrarily-structured electron wave illumination provided by the PELM will be a game-changer in our ability to visualize the dynamic behavior of nanoscale materials and disentangle the interplay of their multiple degrees of freedom. Ultimately, such control will provide a direct handle on the physical and chemical properties of quantum materials, playing a decisive role to address the grand challenges that the world is facing nowadays, especially regarding 'energy', 'information' and 'health'.

**Acknowledgements**
This work is part of the SMART-electron project that has received funding from the European Union's Horizon 2020 Research and Innovation Programme under grant agreement No 964591. Partial additional funding includes European Research Council (Consolidator grant No. 771346 ISCQuM, Advanced Grant No. 789104-eNANO), Spanish MCINN (PID2020-112625GB-I00 and CEX2019-000910-S), Generalitat de Catalunya (CERCA and AGAUR), and Fundaciós Cellex and Mir-Puig.

# METHODS

## Sample preparation
The thin plates used in the experiments are made of a 30-nm thick silver layer deposited via sputtering at a rate of 5.8 Å/s on a 20-nm thick $Si_3N_4$ membrane placed on a silicon support. The plate is mounted on a double-tilt TEM sample holder to ensure full rotation around the transverse $x$ and $y$ axes (see Supplementary Fig. 1). In particular, the plate rotates with an angle $\vartheta$ around the $y$ axis and an angle $\alpha$ around the $x$ axis.

## UTEM experiment
The experiments are performed in an ultrafast transmission electron microscope, whose technical details are described in Ref. 52. Briefly, we illuminate the thin $Ag/Si_3N_4$ film described above using 600-fs infrared pulses ($\hbar\omega = 1.57$ eV photon energy) at a repetition rate of 1 MHz delivered by a KMLabs Wyvern X Ti:Sapph amplified laser system. The light propagates within the $y$–$z$ plane and forms an angle $\delta$ ~ 4.5° with the $z$ axis, as shown in Supplementary Fig. 1. For the experiments presented in the main text, we use a peak light-field amplitude ~ $10^7$ V/m. The delay between electron and photon pulses is varied via a computer-controlled delay line. Simultaneously, a small portion of the infrared laser output is frequency-tripled and directed toward the $LaB_6$ cathode of a JEOL JEM-2100 transmission electron microscope from which femtosecond single-electron pulses are generated via photoemission. The electron pulses are accelerated to an energy $E_0$ = 200 keV along the $z$ axis, and then focused onto the thin $Ag/Si_3N_4$ plate. Inelastically-scattered electrons are then recorded in space and time with a Gatan Imaging Filter (GIF) spectrometer coupled to a K2 direct detector camera for energy-filtered real-space imaging and spectroscopy (~ 1.2 eV energy resolution). The real-space images presented in this work are acquired in energy-filtered mode by selecting a ~ 15 eV window centered on the energy gain side, leaving out the zero-loss peak (ZLP). This procedure directly provides the spatial distribution of the inelastically-scattered electrons interacting with the optical fields at the silver surface of the thin film.

## Spatial light modulation
As shown in Fig. 1a, before reaching the Ag thin film inside the microscope, the infrared light beam is directed toward the surface of a SLM working in reflection mode. The SLM is a PLUTO 2.1 from HOLOEYE Photonics AG, suitably designed to modulate light close to 800 nm wavelength and featuring a high reflectivity and endurance to high light power. The SLM display, which features a resolution of 1920x1080 pixels with 8 μm pixel size, is addressed with phase functions via standard graphics cards as an extended monitor device. To ensure optimal light modulation efficiency, we introduce a polarizer to set the horizontal polarization state of the infrared light beam before reaching the SLM, and illuminate the SLM surface at an angle of ~ 7° with respect to its normal. Following SLM modulation, the light field is then focused on the $Ag/Si_3N_4$ film by means of a plano-convex lens with a focal length of ~ 25 cm. Such configuration makes the silver surface the conjugate plane of the SLM surface. The spatial maps of the light transverse distribution are then measured via an optical beam profiler placed at a distance from the lens equivalent to the Ag surface and following a path that mimics the one travelled by the light beam within the microscope.

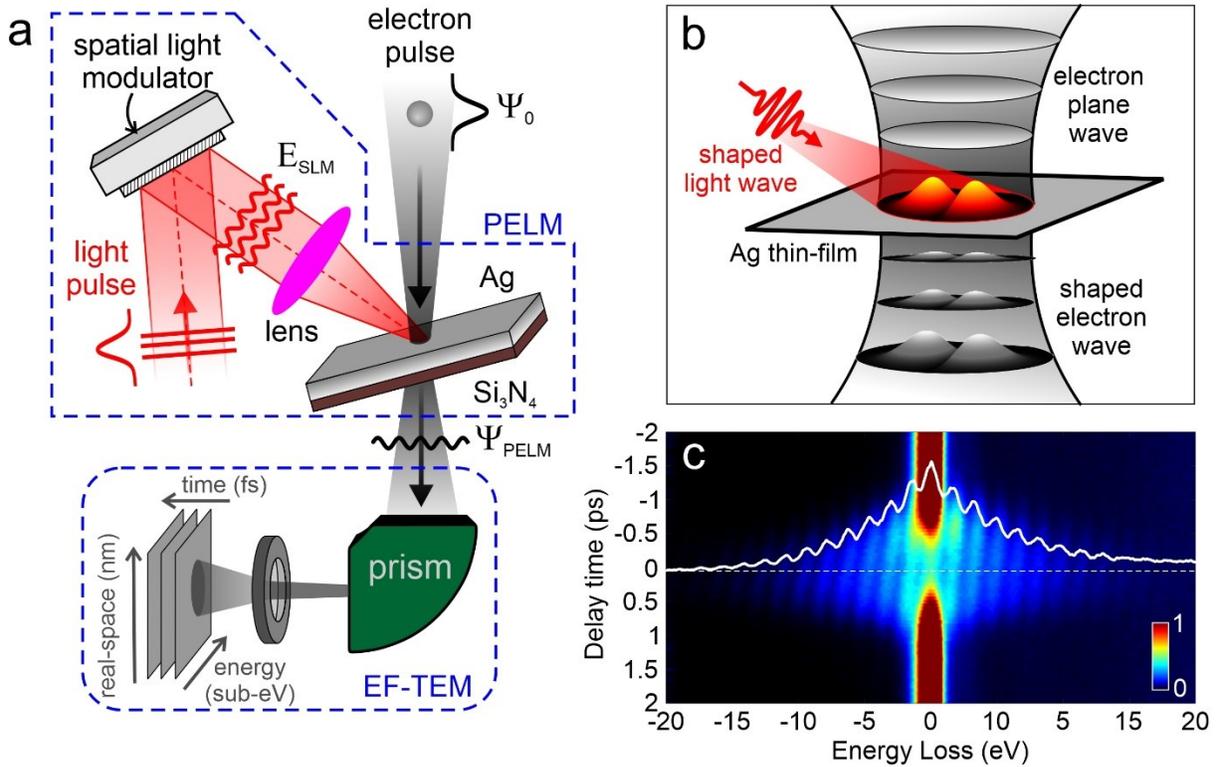

**Figure 1. Schematics of a Photonic free-ELectron Modulator (PELM). (a)** An external spatial light modulator (SLM) is used to imprint an arbitrarily-shaped amplitude and phase pattern on the optical field. The light beam is then focused on a thin Ag/Si$_3$N$_4$ film inside an ultrafast transmission electron microscope (UTEM). The SLM is placed in the conjugate plane with respect to the thin plate. Femtosecond electron pulses in the UTEM impinge on the Ag/Si$_3$N$_4$ film and interact with the modulated femtosecond light-pulse field at the Ag surface *via* stimulated inverse transition radiation. The inelastically-scattered electrons are then imaged in space, time and energy by means of electron energy-loss spectroscopy (EELS) performed in our EF-TEM setup. **(b)** Schematic picture of the optical modulation of a free-electron plane wave by a shaped light wave with a Hermite-Gaussian transverse profile. **(c)** Sequence of measured EELS spectra (color map) plotted as a function of the delay time between the electron and light pulses. Sidebands at energies $\pm \ell \hbar \omega$ (in our case $\hbar \omega = 1.57\ \mathrm{eV}$) relative to the zero-loss peak (ZLP) are visible, where $\ell$ is the net number of exchanged photons. White solid line: selected EELS spectrum measured at t = 0 ps, corresponding to the temporal and spatial coincidence between electron and light pulses.

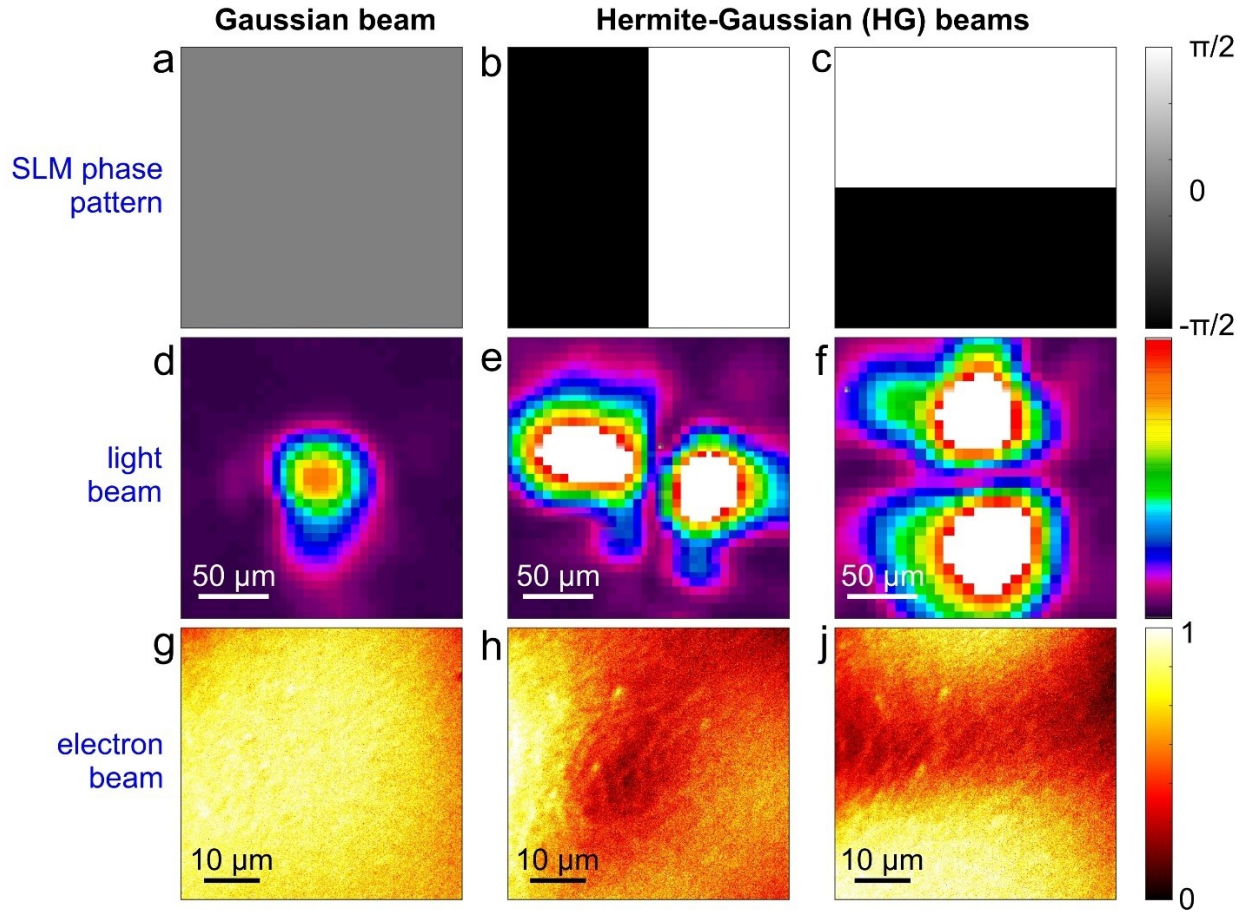

**Figure 2. Experimental demonstration of transverse optical manipulation of free electrons via arbitrarily-shaped ultrafast light fields. (a)-(c)** Phase patterns implemented on the SLM used to modulate the light field: (a) homogeneous phase distribution, (b) π phase shift along the horizontal direction, and (c) π phase shift along the vertical direction. **(d)-(f)** Light transverse profiles measured for the corresponding SLM phase patterns in (a)-(c): a Gaussian profile in (d) a two-lobed horizontal Hermite-Gaussian profile ($HG_{10}$) in (e) and a two-lobed vertical Hermite-Gaussian profile ($HG_{01}$) in (f). **(g)-(j)** Inelastically-scattered electron spatial maps measured under optical illumination with (g) Gaussian and (h)-(j) Hermite-Gaussian profiles, obtained by the SLM modulation of the ultrafast light pulses shown in (a)-(c). All images are taken when electron and light wavepackets have maximum temporal overlap.

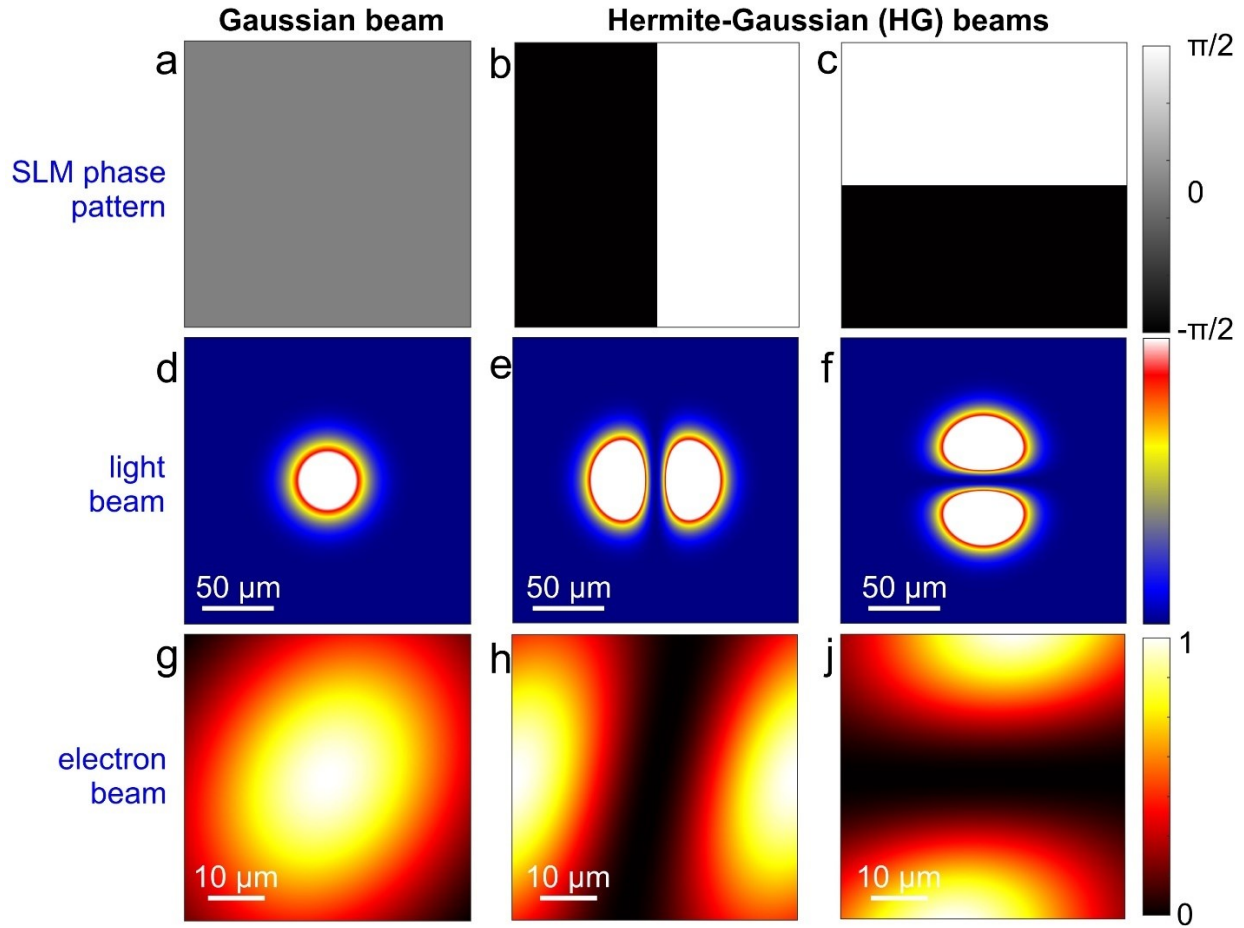

**Figure 3. Theoretical calculations of transverse optical manipulation of free electrons via arbitrarily shaped ultrafast light fields.** We present simulations corresponding to the same plots and labels as in Fig. 2. The asymmetry in the electron beam profiles in (g)-(j) is due to the tilt angle of the incident light direction relative to the electron beam

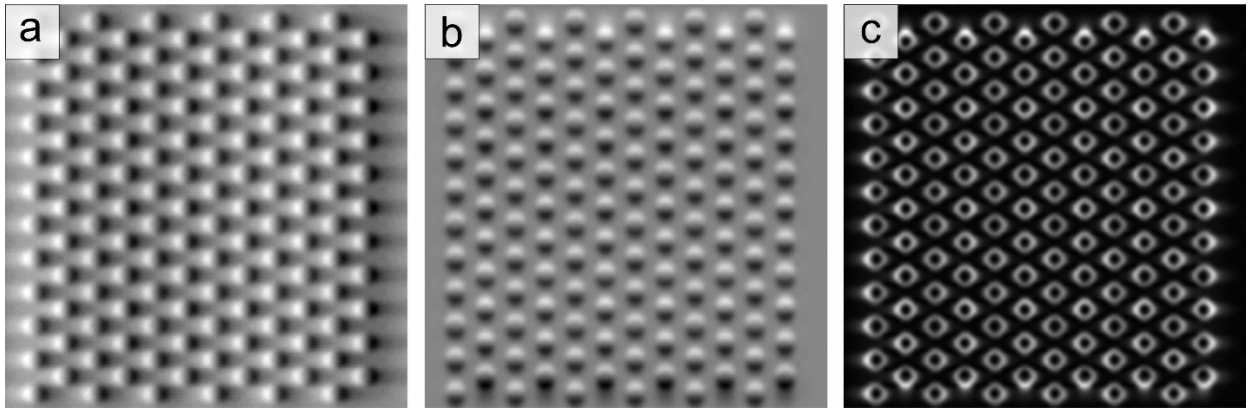

**Figure 4. Enhancing topographical contrast of a weak-phase object using Hermite-Gaussian electron beams. (a,b)** Simulated TEM images of an ordered array of Skyrmions (diameter of ~ 50 nm, separation of 100 nm) obtained at the focus condition (zero defocus) with shaped electrons having (a) $HG_{10}$ and (b) $HG_{01}$ symmetries. The Hermite-Gaussian modulation of the electron wavepacket induced by the PELM is equivalent to the application of a Hilbert phase plate with the corresponding symmetry. **(c)** Quadratic average of the two images reported in panels (a) and (b), showing an enhanced local contrast of the atomic array of Skyrmions. The field of view in all panels is 1.13 x 1.13 $\mu m^2$.

# Supplementary Information

# Ultrafast Transverse Modulation of Free Electrons by Interaction with Shaped Optical Fields

I. Madan[1], V. Leccese[1], A. Mazur[2], F. Barantani[1,3], T. La Grange[1], A. Sapozhnik[1], S. Gargiulo[1], E. Rotunno[4], J.-C. Olaya[2], I. Kaminer[5], V. Grillo[4], F. J. García de Abajo[6,7], F. Carbone[1,*], and G. M. Vanacore[8,*]

1. Institute of Physics, École Polytechnique Fédérale de Lausanne, Lausanne, 1015, Switzerland
2. HOLOEYE Photonics AG, Volmerstrasse 1, 12489 Berlin, Germany
3. Department of Quantum Matter Physics, University of Geneva, 1211 Geneva, Switzerland
4. Centro S3, Istituto di Nanoscienze-CNR, 41125 Modena, Italy
5. Department of Electrical and Computer Engineering, Technion, Haifa 32000, Israel
6. ICFO-Institut de Ciencies Fotoniques, The Barcelona Institute of Science and Technology, 08860 Castelldefels (Barcelona), Spain
7. ICREA-Institució Catalana de Recerca i Estudis Avançats, Passeig Lluís Companys 23, 08010 Barcelona, Spain
8. Department of Materials Science, University of Milano-Bicocca, Via Cozzi 55, 20126 Milano, Italy

*To whom correspondence should be addressed: giovanni.vanacore@unimib.it, fabrizio.carbone@epfl.ch

# SUPPLEMENTARY SECTION 1:
# Theoretical simulation of electron-light interaction

With a reflectivity larger than 99%, a metal skin depth of ~ 11 nm – smaller than the silver layer thickness of ~ 30 nm –, and the large permittivity at the photon energy of 1.57 eV, the performance of the Ag/Si$_3$N$_4$ film used in the experiments is close to that of a perfect mirror. In this approximation, the electron-light coupling β parameter is given by Eq (3) in the main text. To compute such equation, we need to define the incident and reflected light electric fields and wave vectors for the tilting geometry adopted in the experiments.

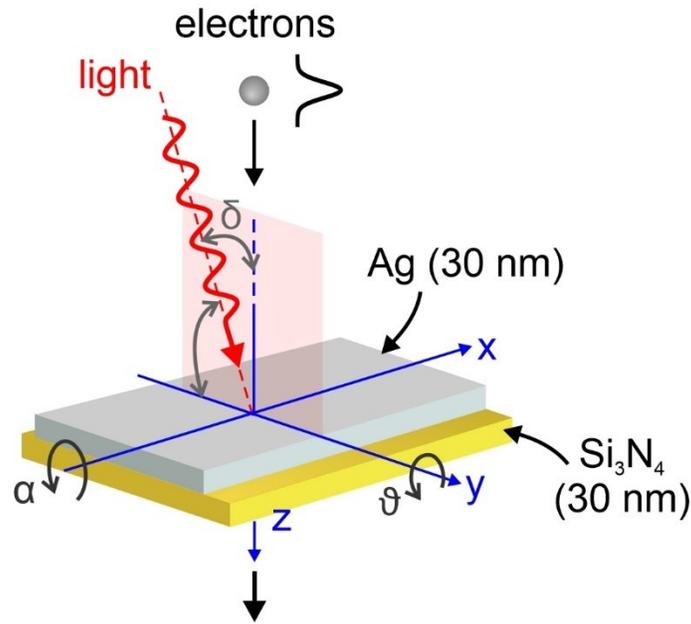

**Supplementary Figure 1. Schematic diagram of the experimental geometry.** A Ag/Si$_3$N$_4$ plate is mounted on a double-tilt TEM sample holder to ensure full rotation around the transverse x and y axes. In particular, the plate rotates by an angle $\vartheta$ around the y axis and by an angle $\alpha$ around the x axis. The electron beam moves along the z axis, whereas the light beam propagates within the y–z plane and forms an angle $\delta$ ~ 4.5° with respect to the z axis.

Taking Supplementary Fig. 1 as reference, we denote with $[x, y, z]$ the coordinates in the unrotated frame; $[x', y', z']$ in the frame after rotating by a tilt angle $\vartheta$; and $[x'', y'', z'']$ in the frame after a subsequent rotation by an angle $\alpha$. These two rotations are described by the relations:

$$x' = x\cos\vartheta + z\sin\vartheta;\ y' = y;\ z' = -x\sin\vartheta + z\cos\vartheta \qquad (S1)$$

and

$$x'' = x';\ y'' = y'\cos\alpha - z'\sin\alpha;\ z'' = y'\sin\alpha + z'\cos\alpha. \qquad (S2)$$

Combining Eq. (S1) and (S2), we find

$$x'' = x\cos\vartheta + z\sin\vartheta; \quad y'' = y\cos\alpha - (z\cos\vartheta - x\sin\vartheta)\sin\alpha,$$
$$z'' = y\sin\alpha + (z\cos\vartheta - x\sin\vartheta)\cos\alpha. \tag{S3}$$

The incident light electric field and wave vector in the unrotated frame can be written as

$$\mathcal{E}^{inc}(x,y) = A\mathcal{E}_0(x,y) \begin{bmatrix} 0 \\ \cos\delta \\ \sin\delta \end{bmatrix},$$

$$k^{inc} = k_0 \begin{bmatrix} 0 \\ \sin\delta \\ \cos\delta \end{bmatrix}. \tag{S4}$$

By applying the transformation defined by Eq. (S3), we obtain

$$\mathcal{E}^{inc''}(x'',y'') = A\mathcal{E}_0''(x'',y'') \begin{bmatrix} \sin\delta\sin\vartheta \\ \cos\delta\cos\alpha - \sin\delta\cos\vartheta\sin\alpha \\ \cos\delta\sin\alpha + \sin\delta\cos\vartheta\cos\alpha \end{bmatrix},$$

$$k^{inc''} = k_0 \begin{bmatrix} \cos\delta\sin\vartheta \\ \sin\delta\cos\alpha - \cos\delta\cos\vartheta\sin\alpha \\ \sin\delta\sin\alpha + \cos\delta\cos\vartheta\cos\alpha \end{bmatrix}. \tag{S5}$$

Following the reflection from the (ideal) silver surface, the x and y components of the reflected quantities are the opposite of the x and y components of the incident quantities, while the z component remains the same. The reflected light electric field and wave vector are thus given by

$$\mathcal{E}^{ref''}(x'',y'') = A\mathcal{E}_0''(x'',y'') \begin{bmatrix} -\sin\delta\sin\vartheta \\ -\cos\delta\cos\alpha + \sin\delta\cos\vartheta\sin\alpha \\ \cos\delta\sin\alpha + \sin\delta\cos\vartheta\cos\alpha \end{bmatrix},$$

$$k^{ref''} = k_0 \begin{bmatrix} -\cos\delta\sin\vartheta \\ -\sin\delta\cos\alpha + \cos\delta\cos\vartheta\sin\alpha \\ \sin\delta\sin\alpha + \cos\delta\cos\vartheta\cos\alpha \end{bmatrix}. \tag{S6}$$

Considering that the unit vector along the z axis, which we denote as $u_z$, is transformed from [0,0,1] in the unrotated framework to $u_z'' = [\sin\vartheta, -\cos\vartheta\sin\alpha, \cos\vartheta\cos\alpha]$ following α and ϑ rotations, we can calculate the projection of the incident/reflected electric field and wave vector in the rotated framework as

$$\mathcal{E}_z^{inc''} = \mathcal{E}^{inc''} \cdot u_z'' = A\mathcal{E}_0''\sin\delta,$$
$$k_z^{inc''} = k^{inc''} \cdot u_z'' = k_0\cos\delta, \tag{S7}$$

$$\begin{aligned}\mathcal{E}_z^{ref''} &= \mathcal{E}^{ref''}\cdot u_z'' = A\mathcal{E}_0''[\sin\delta(\cos^2\vartheta\cos2\alpha - \sin^2\vartheta) + \cos\delta\cos\vartheta\sin2\alpha],\\ k_z^{ref''} &= k^{ref''}\cdot u_z'' = k_0[\cos\delta(\cos^2\vartheta\cos2\alpha - \sin^2\vartheta) + \sin\delta\cos\vartheta\sin2\alpha].\end{aligned} \quad (S8)$$

Finally, the transverse light profile in the unrotated framework is given by either a Gaussian distribution or an Hermite-Gaussian distribution according to the settings used for the SLM:

$$\begin{aligned}\mathcal{E}_0^G(x,y) &= \exp\left[-\frac{(x^2+y^2)}{2\sigma_L^2}\right],\\ \mathcal{E}_0^{HG10}(x,y) &= \exp\left[-\frac{(x^2+y^2)}{2\sigma_L^2}\right]\frac{2x}{\sigma_L},\\ \mathcal{E}_0^{HG01}(x,y) &= \exp\left[-\frac{(x^2+y^2)}{2\sigma_L^2}\right]\frac{2y}{\sigma_L},\end{aligned} \quad (S9)$$

where $2\sqrt{2ln2}\sigma_L$ defines the full width at half maximum (FWHM) of the laser beam. By applying the transformation defined by Eq. (S3) with $z = 0$, we can retrieve the transverse light profiles in the rotated framework:

$$\begin{aligned}\mathcal{E}_0^{G''}(x'',y'') &= \exp\left[-\frac{(\Gamma_1 x''^2 + \Gamma_2 y''^2 - 2\Gamma_3 x''y'')}{2\sigma_L^2}\right],\\ \mathcal{E}_0^{HG10''}(x'',y'') &= \exp\left[-\frac{(\Gamma_1 x''^2 + \Gamma_2 y''^2 - 2\Gamma_3 x''y'')}{2\sigma_L^2}\right]\frac{2x''}{\cos\vartheta\sigma_L},\\ \mathcal{E}_0^{HG01''}(x'',y'') &= \exp\left[-\frac{(\Gamma_1 x''^2 + \Gamma_2 y''^2 - 2\Gamma_3 x''y'')}{2\sigma_L^2}\right]\frac{2}{\sigma_L}\left(\frac{y''}{\cos\alpha} - \tan\vartheta\tan\alpha x''\right),\end{aligned} \quad (S10)$$

where

$$\begin{aligned}\Gamma_1 &= \frac{1}{\cos^2\vartheta} + \tan^2\vartheta\tan^2\alpha,\\ \Gamma_2 &= \frac{1}{\cos^2\alpha},\\ \Gamma_3 &= \frac{\tan\vartheta\tan\alpha}{\cos\alpha}.\end{aligned} \quad (S11)$$

Calculations of the inelastically-scattered electron spatial maps are then implemented by plugging the expressions found in Eqs. (S7), (S8), (S10) and (S11) within Eqs. (3) and (4) in the main text, using the following geometrical and experimental parameters: $\delta = 4.5°$, $\alpha = 12.7°$, $\vartheta = 35°$, $\sigma_L = 23.3$ µm, $v = 0.7c$, and $A = 10^7$ V/m.